\documentclass{article}
\usepackage{graphicx}

\begin{document}

\title{Modified SIMD architecture suitable for single-chip implementation}

\author{Junichiro Makino\\
Department of Astronomy, School of Science, The University of Tokyo, 
 Tokyo 133-0033\\
makino@astron.s.u-tokyo.ac.jp}

\maketitle

\begin{abstract}

We describe a modified SIMD architecture suitable for single-chip
integration of a large number of processing elements, such as 1,000
or more. Important differences from traditional SIMD designs are:

\begin{description}

\item{a)} The size of the memory per processing elements is kept small.

\item{b)} The processors are organized into groups, each with a small
buffer memory. Reduction operation over the groups is done in hardware.

\end{description}

The first change allows us to integrate a very large number of
processing elements into a single chip.  The second change allows us
to achieve a close-to-peak performance for many scientific applications
like particle-based simulations and dense-matrix operations.

\end{abstract}

\section{Introduction}

SIMD parallel processing is a very old idea, with several successful
implementations such as Illiac-IV\cite{Hord1982}, ICL/AMT DAP, Goodyear MPP\cite{Potter1985}, TMC CM-1/2\cite{Hillis1985}, and
INFN/Quadrics APE-100 and APEmille. These are large machines made of
up to 64K processors, each with its own local memory. Except for the
APE machines which were designed for LQCD calculation, all of these
machines were built before 1990. 

There is another form of SIMD architecture. Almost all
recent microprocessors have some form of SIMD processing units, with 4 or
more arithmetic units. These include VIS of SPARC, Altivec of PowerPC,
MVI of Alpha, 3DNow! of AMD and MMX, SSE, and SSE2 of Intel x86.

Though both of these two architectures are called "SIMD", the actual
hardware implementation and programming models are completely
different. In the former case of large-scale SIMD parallel machines,
each  processing element has its own memory and the address generation
unit for it, and they are connected through some routing network. In the
latter case, the instruction set of a single processor is extended to
handle a single long word as a vector of multiple short words. Thus,
essentially only the arithmetic units are duplicated, and they are
connected to single memory unit (or single L1 cache) through a single
datapath.

The former architecture is for large machines made of a
number of processing chips, each with one or few processors. The
latter is for a single processor chip. With present-day
technology we could in principle integrate thousands or more of
processors used in machines like CM-1 into a single chip, and yet what
is currently used is SIMD extensions with just a few arithmetic
units. 

In this paper, we describe a new architecture with which we can
integrate a very large number of processors into a single chip and
achieve reasonable efficiency for a wide range of applications.

\section{Greatly Reduced Array of Processor Elements}

\label{sect:basic}

\begin{figure}
\begin{center}
\leavevmode
\includegraphics[width=8 cm]{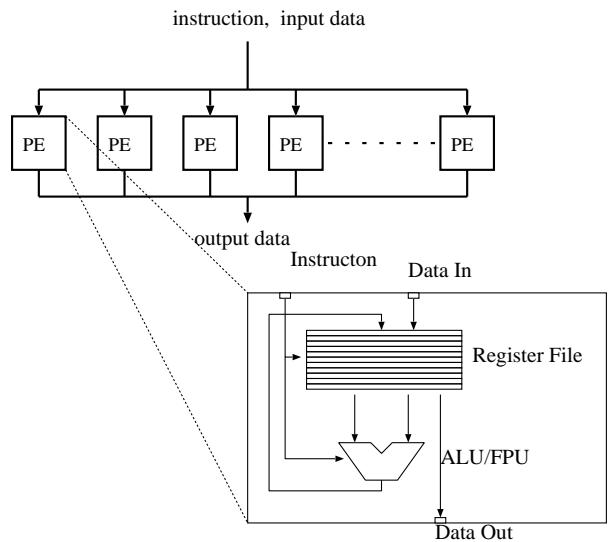}
\end{center}
\caption{Basic structure of an SIMD processor}
\label{fig:basicsimd}
\end{figure}

Figure \ref{fig:basicsimd} shows the basic architecture we discuss in
this paper. It consists of a number of processing elements, each of
which consists of an FPU and a register file. They all receive the
same instruction from outside the chip, and perform the same
operation.

Compared to the classic SIMD architecture such as that of TMC CM-2, the main
difference are the followings.

\begin{description}

\item{a)} PEs do not have large local memories.
\item{b)} There is no communication network between PEs.

\end{description}

We introduce these two simplifications so that a large number of PEs
can be integrated into a single chip. If we want to have a memory with
meaningful size connected to each PE, the only economical way is to
attach DRAM chips. A chip which integrates logics and DRAM memory can
in principle be made, but the price per bit of the DRAM memory of such
a chip is about two orders of magnitude higher compared to
that of commercial DRAM chips. However, once we decide to use external
memory, it becomes very difficult to integrate large number of
processors into a chip. Consider an example of a chip with 100
processors, each with single arithmetic unit. If the clock frequency
is 1 GHz, the peak speed of the chip is 100 Gflops. If we want to add
the memory units which can supply one word per clock cycle to these
100 processors, we need the memory bandwidth of 800 GB/s, or around
100 times more than that of the latest microprocessors. Clearly, it is
not easy to achieve such a high memory bandwidth.
If we eliminate the local memory of processors, we can integrate very
large number of processors into a chip.

A communication network, as far as it is limited into a single chip, is
not very expensive. A two-dimensional mesh network would be quite
natural, for physically two-dimensional array of PEs on a single
silicon chip. However, such a two-dimensional network poses a very hard
problem, if we try to extend it to multi-chip systems.
With current and near-future VLSI technology, it is possible to
integrate more than 1000 PEs to a single chip, each with fully
pipelined FPUs. Thus, a 2D array will have the dimensions of 32 by 32,
and the minimum number of external links necessary to construct a 2D
network is $32\times 4 = 128$. If we want to have, say, 16 wires per
link, the total number of pins necessary is 2,048. To make such a
large number of pins work with reasonable data rate is not impossible,
but very costly.

If we eliminate the inter-PE communication network right from the
beginning, we have no problem in constructing multi-chip systems,
since PEs in different chips need not be connected.

Thus, this simple architecture has two advantages. First, we can
integrate a very large number of PEs into a single chip. Second, a
system with multiple chips is easy to construct. As a result, peak
performance of the system can be very high.

Important question here is if any real application can actually take
advantage of this architecture.  We consider several examples and
extend the architecture in the next section.

Note that this SIMD processor works as an attached processor to
general-purpose CPU. Since the on-chip memory is limited to just the
register files of PEs, the SIMD processor itself cannot run any
application which requires large amount of memory. Thus, we need to
move only the part of the application program which can be efficiently
done on the SIMD processor. This of course means there will be
communication overhead. 

Before we proceed to the next section, we need a name for the proposed
architecture. Since the main difference between the proposed
architecture and previous SIMD architecture is the removal of elements
like local memory and inter-PE network, we call this architecture
Greatly Reduced Array of Processor Elements, or GRAPE. The similarity
of this name to the GRAPE for astronomical $N$-body
simulations\cite{Sugimoto1990} is a pure coincidence. 

\section{Modification to the basic architecture}

\subsection{Particle-based simulations}

In many particle-based simulations such as classical MD or
astrophysical $N$-body simulations,  the most expensive part of
calculation is the evaluation of particle-particle interactions. In
general, it has the form
\begin{equation}
f_i = \sum_{j\ne i} g(x_i, x_j),
\label{eq:pp}
\end{equation}
where $x_i$ denotes the variables associated with particle $i$,
$g(x_i, x_j)$ is some generalized ``force'' from particle $j$ to
particle $i$, and $f_i$ is the total ``force'' on particle $i$. At least
formally, the summation is taken over all particles in the system.
Therefore the calculation cost is $O(N^2)$, where $N$ is the total number of
particles in the system. In some cases the interaction is
of short-range nature and we can apply some cut-off length. If the
interaction is long-ranged, we might be able to use approximate
algorithms such as FMM\cite{GreengardRokhlin1987} or Barnes-Hut
tree\cite{BarnesHut1986}.

In these cases, however, the most expensive part is still the direct
evaluation of equation (\ref{eq:pp}). The basic SIMD structure we
discussed in the previous section is quite suited to calculations of
this type. In the simplest case, we first load data of particles on
which we calculate the interaction to the registers of PEs. In other
words, we first write one $x_i$ to each PE.  Then we broadcast one
$x_j$ to all PEs and let then calculate the force from this particle
$j$ to their particles. We repeat sending particles $x_j$ until all
particles are sent, and then read the calculated results $f_i$ in
each PE. Remaining calculations such as the time integration of
particles are done on the host computer which controls the SIMD
processor.

In this way, the size of the system is not limited by the size of the
SIMD processor array. Also, since the program, except for that for the
evaluation of particle-particle interaction, runs completely on the
host computer, it is relatively easy to adopt existing programs to
take advantage of the array processor.

If the number of particles is much smaller than
the number of PEs, the efficiency would become low. Even when the total
number of the particles is large, if the interaction is short-ranged,
the number of particles with which one particle physically interact
can be much smaller than the number of PEs.

\begin{figure}
\begin{center}
\leavevmode
\includegraphics[width=12 cm]{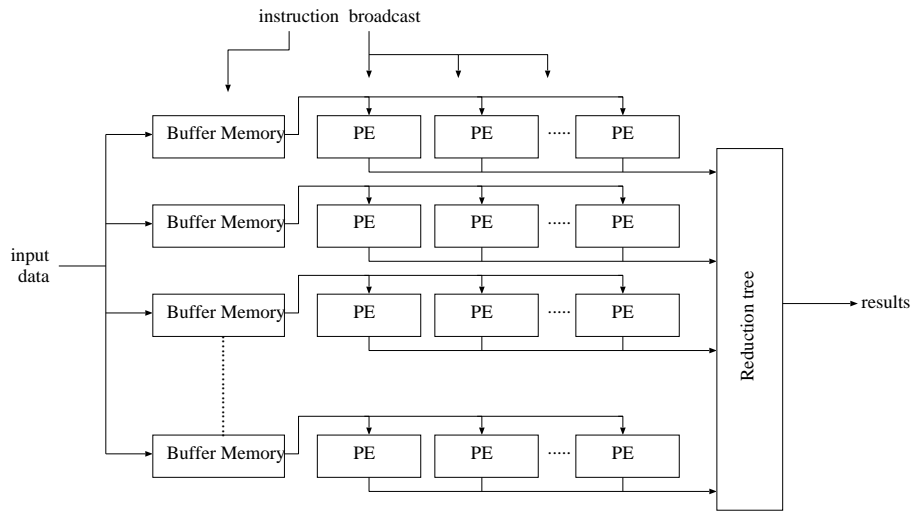}
\end{center}
\caption{Modified SIMD architecture}
\label{fig:ijparallelsimd}
\end{figure}

This problem can be solved in many different ways. One possibility is
to organize the processors into groups, as shown in figure \ref{fig:ijparallelsimd}.

In this modified architecture, PEs are organized into groups, each
with small buffer memory. These groups are connected to a reduction
network. The host computer can  either write data to individual buffer
memories or broadcast the same data to all buffer memories. In this
way, PEs in different groups can calculate the forces from different
particles. In addition, the reduction network allows multiple PEs in
different groups to calculate the force on the same particle from
different particles. Thus, the efficiency for small-$N$ systems or
short-range force is greatly improved.

Note that the hardware cost of the buffer memory and reduction network
is very small, since their cost is proportional to the number of
groups, which is a small fraction of the number of PEs.

\subsection{Dense Matrix operations}

For any dense-matrix algorithms, the basic operation is matrix
multiplication $C=AB$. Thus, the key question is if our proposed architecture
can achieve reasonable performance for matrix multiplication.

We consider the modified architecture discussed in the previous
section. In this architecture, it is easy to implement parallel matrix
multiplication. The basic idea is to block-subdivide the matrix $A$
into two dimensions in the same way as in the  standard Canon's algorithm and
load them to each PE. Then we take one column of $B$ and divide it to
$n$ pieces, where $n$ is the number of groups, and send these pieces to
the buffer memories of group. We then calculate $c=Ab$ on each PE on
each group. By taking summation over groups, we obtain one row of $C$.

\subsection{Greatly Reduced Array of Processor Elements with Data
Reduction}

We have changed the basic SIMD architecture by grouping PEs and adding
the buffer memories and the reduction network. Therefore, we call this
architecture GRAPE-DR, or Greatly Reduced Array of Processor Elements with Data
Reduction.

\section{ Discussions and Summary}

Modern GPUs and some DSPs have the SIMD architecture very similar to
the basic architecture we described in section \ref{sect:basic}. Most
of them, however, are designed to perform a fixed (though relatively
large) number of operations per data. For example, many DSPs are
designed to perform, say, multiple independent 1K-point FFT operations
in parallel. Thus, the ratio between the calculation speed and the
external memory bandwidth must be kept constant, and the peak
performance of these systems are generally limited by the available
memory bandwidth. In our proposed architecture, we do not intend to run
the applications which require large memory bandwidth, and thus the
peak performance is not limited by the memory bandwidth.

In this paper, we describe a modified SIMD architecture suitable both
for a single-chip implementation and for large systems made of a
number of such chips. It consists of a number of extremely simple yet
fully programmable processors, connected through hierarchical
broadcast/reduction network. In a sense, it is similar to SIMD
parallel processors, but unlike the previous SIMD machines, the local
memory of PE is reduced to small amount, and only the most
compute-intensive part of the application will be run on the SIMD
processor array. A machine base on this architecture is currently
under development\cite{GDR}.

\section*{Acknowledgments}

The author thanks Kei Hiraki and Mary Inaba for helpful discussions and
useful comments on the manuscript. He also thanks Toshiyuki Fukushige,
Yoko Funato, Piet Hut, Toshikazu Ebisuzaki, and Makoto Taiji for
discussions related to this work.  This research is partially
supported by the Special Coordination Fund for Promoting Science and
Technology (GRAPE-DR project), Ministry of Education, Culture, Sports,
Science and Technology, Japan.

\def\MN{MN}


\newcommand{\etalchar}[1]{$^{#1}$}
\newcommand{\noopsort}[1]{} \newcommand{\printfirst}[2]{#1}
  \newcommand{\singleletter}[1]{#1} \newcommand{\switchargs}[2]{#2#1}

\end{document}